\documentclass[epj]{svjour}
\usepackage{graphics}
\usepackage{amsmath}
\usepackage{amssymb}
\begin{document}
\title{
Possibility of T-violating P-conserving magnetism and its
contribution to the
T-odd P-even neutron-nucleus
forward elastic scattering amplitude.
}
\author{Sergey L. Cherkas }
\institute{Institute of Nuclear Problems, 220050 Minsk, Belarus}
\date{Received:29 August 2000/ Revised version:22 January 2001}
%
\abstract{
T-violating P-even magnetism is considered.
The magnetism arises from the T-violating P-conserving
vertex of
a spin 1/2 particle
 interaction  with the electromagnetic
field.
The vertex varnishes for a particle on the mass shell.
Considering the particle interaction with a point electric charge
we have obtained the T-violating P-even spin dependent potential which is
inversely proportional
to the cubed distance from the charge.
The matrix element of
this potential is zero for particle states on the mass shell,
nevertheless, the potential contributes to the T-odd P-even neutron
forward elastic scattering amplitude
by a deformed nucleus with spin   $S>1/2$.
The contribution arises if we take into account
incident neutron plane
wave distortion by the strong neutron interaction with the
nucleus.
} 
\titlerunning{T-violating P-conserving magnetism and
T-odd neutron-nucleus
forward elastic scattering }
\maketitle
\section{Introduction}
\label{intro}
In connection with the direct observation of
time-reversal symmetry
violation in
the system  of $K^0-\bar K^0 $ mesons
\cite {cplear}
it would be interesting to detect
T-violation in other nuclear or atomic systems.
However, the Standard Model predicts very small T-violating effects in 
nuclear and
atomic physics, so we are forced to search for new interactions.
It is necessary to distinguish a P- T- odd interaction from
a P-even T-odd one. While
there are rather rigid restrictions on the
strength constants of the first type interactions, obtained from  dipole
moment 
measurements
of atoms and particles, restrictions on
the constants of P-even T-odd interactions
are not so strong.
As is known
the null
 test for
the latter kind of interaction
is the observation  of a
$ \sim (\boldsymbol \sigma\times \vec k
\cdot \vec S) (\vec k
\cdot \vec S) $  five fold  correlation term
in the forward elastic scattering amplitude of
a spin $1/2 $ particle
by a particle with a spin $S\ge 1$   \cite{bar1,conzet,beyer,cher},
where $ \vec k $ is
the incident particle momentum,
$\boldsymbol \sigma$ is the Pauli matrix of an incident
particle and $\vec S$ is the nucleus spin operator.
The relevant
experiments
have been
carried out \cite {exp,huf0} for a $^{165}\mbox{Ho}$ target  and
now are
planed to be performed on a super-conducting synchrotron COSY
\cite {COSY} with deuteron. Usually 
the
P-conserving breakdown of the time reversal
symmetry is considered on the basis of the $\rho$ and $A_1$ meson
 Lagrangian \cite{sim}.
In this paper we consider
another phenomenological possibility, namely, T-violating P-conserving
magnetism and its contribution
 to the aforementioned five fold correlation.
\section{Long-range
T-non-invariant P-even electromagnetic interaction.}
\label{sec:1}
\begin{figure}
\vspace{0.cm}
\hspace{1.5 cm}
\resizebox{0.25\textwidth}{!}{
  \includegraphics{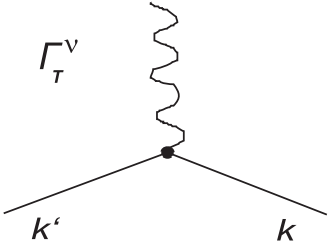}
}
\vspace{1.cm}       
\caption{T-odd P-even vertex of the particle interaction with the electromagnetic field.}
\label{fig:1}       
\end{figure}

The magnetism
can be introduced
 by the T-violating P-conserving
vertex function of
a spin 1/2 particle 
interaction  with the electromagnetic
field \cite{t1,haff}:
\begin {equation}
\Gamma^\eta_T=\mu_T\frac{\mbox{i}}{2\,m^3}
(Pq)\sigma^{\eta\nu}q_\nu~,
\label{vert}
\end {equation}
where $P=k^\prime+k $, $q=k^\prime-k$  (Fig. \ref{fig:1}),
$m$ is the particle mass and
$\sigma^{\eta\nu}=\frac{\gamma^\eta\gamma^\nu-\gamma^\nu\gamma^\eta}{2}$.
Let us consider T-odd scattering of a particle by a point
electric charge $Ze$.

After the application of
ordinary  diagram techniques \cite{ber}
we obtain the appropriate matrix element corresponding to the
diagram in Fig. \ref{fig:1}:
\begin{eqnarray}
{ \mathcal M} = -\mu_T\frac {\mbox{i}e} {2\,m^3}
(Pq)
\bar u(k^\prime)
 \sigma ^ {0\nu}  q_\nu u(k)
{\mathcal A}^{(e)}_0~,
\label{matrx}
\end{eqnarray}
where
${\mathcal A}^{(e)}_0(q)=\frac{4\,\pi Z e}{\vec q^2}$ is
the
Fourier transform of the
Coulomb potential of the electric charge and $u(k)$ is
the
particle bispinor.
Setting $q = (0, \vec q) $ and substituting
$$u (k) = \left (\begin {array}{c}
 \sqrt {\varepsilon+m} \; \phi \\
(\varepsilon+m)^{-1/2} \; (\boldsymbol \sigma \vec k) \; \phi
 \end{array}
\right) ~
$$
($ \phi $ is the spin wave function of a particle
and $\varepsilon$ is the particle energy including its rest mass)
into (\ref{matrx}) we find the
T-odd P-even
scattering amplitude of a particle by a point electric charge for
small transferred momentum $ \vec q $:
\begin{equation}{\mathfrak f}(\vec q) =
\frac{\mathcal M}{4\pi} = - \mu_T\frac{2Ze^2}{m^3} \frac {(\vec k
\vec q) (\boldsymbol \sigma\times \vec k\cdot \vec q)} {\vec q^2}.
\label{f}
\end{equation}
While evaluating
 the scattering amplitude we
consider a particle to be on the  mass shell
$(\vec k^2=(\vec k+\vec q)^2=\varepsilon^2-m^2)$
everywhere
except for the term
$ (\vec k \vec q) $. If the
particle is completely on the mass shell
$ (\vec k \vec q) =0$ and the amplitude (\ref{f}) vanishes. The
dependence of the amplitude on the 
transferred  momentum
$ \vec q $ looks like
that for the
magnetic dipoles scattering
amplitude. So, it turns out
that the interaction is
long-range.
In  \cite{haff}
the  conclusion 
(repeated in
the monograph of \cite {blin})
had been drawn of the non-existence of
a
long-range T-odd P-even potential
(i.e. it is decreasing as $1/r^3$ or weaker with distance
\cite {Dau}
).
 However, we'll see that this conclusion
does not concern of-mass-shell potentials.

We can consider the particle scattering
in the framework of the Schroedinger equation with
relativistic mass \cite {ber,for}:
\begin {equation}
( \nabla^2 + \vec k^2) \Psi ({\vec r})
=2\varepsilon V (\vec r) \Psi (\vec r),
\label{sr}
\end {equation}
which allows us below to take
into
account  incident particle wave
distortion by the strong nucleus interaction.
In the first  Born approximation  the amplitude (\ref {f})
can be obtained from the T-odd
energy dependent
interaction:
\begin{eqnarray}
{V}_T (\boldsymbol r) =- \mu_T\
\frac {3Ze^2} {2\varepsilon m^3}
  \bigl ((
\hat {\vec p} \vec r) \frac {1} {r^5} (
\vec r\cdot\boldsymbol\sigma\times \hat {\vec p})
\nonumber
\\
 + (
\boldsymbol \sigma\times
\hat {\vec p} \cdot \vec r)
\frac {1} {r^5} (\vec r
\hat {\vec p}) \bigr) ~.
\end{eqnarray}
It can be represented by:
\begin {eqnarray}
{V} _T (\vec r) =\mu_T \frac {e} {2 \varepsilon m^3}
\bigl (
\hat {\vec
p} \cdot \{\vec \nabla\otimes \vec E (\vec r) \}
\nonumber
\\
 \cdot (\boldsymbol \sigma
\times \hat
{\vec p}) +
( \boldsymbol \sigma\times \hat {\vec p}) \cdot \{\vec \nabla\otimes \vec E (\vec r)
\} \cdot \hat {\vec p}
\bigr),
\label {telectr}
\end {eqnarray}
where $ \vec E (\vec r) = -\vec \nabla \Phi (\vec r) =Ze {\displaystyle \frac
{\vec r} {r^3}} $ is the strength of the electric field created
by a charge at the
point $ \vec r $, the gradient acts on the $\vec E(\vec r)$ only,  
$\hat {\vec p}$ is
the particle momentum operator and $ \otimes $ denotes a direct vector product.
When considering a particle
moving along classical trajectory,
 we should replace the momentum operator by its
classical value.
\begin{figure}
\vspace{0.cm}
\hspace{0.5 cm}
\resizebox{0.4\textwidth}{!}{
  \includegraphics{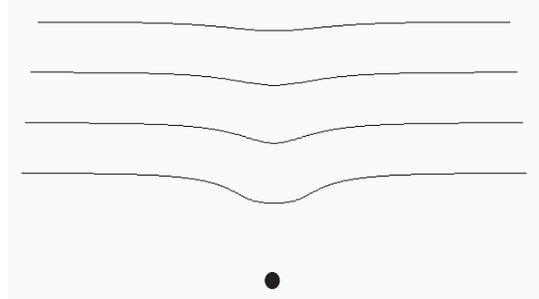}
}
\vspace{-0.5cm}       
\caption{Scematic picture of
classical particle trajectories
(in the first order in interaction)
calculated with  the classical
Hamiltonian obtained from the of-the-mass-shell
interaction through replacing particle momentum operator by
its classical value. }
\label{traj}       
\end{figure}
In the first order in interaction
the trajectories of a
classical particle deflect from a straight line only  in the
vicinity of a scatter  (Fig. \ref{traj}). In the presence of  some
ordinary on-mass-shell interaction, for instance strong one,
the of-mass-shell
T-odd interaction
decreases or increases the  stream of particles in the area of
strong interaction and, thereby, gives a T-odd
contribution to the scattering amplitude.

So, we can see that a moving particle can interact with
a non-uniform electric field
by means of the time reversal violating parity conserving interaction.
\section{
T-odd scattering of a neutron by a deformed
nucleus
with spin $S\ge 1 $ 
}
\label{sec:2}
Let us consider T-odd neutron scattering
 by  a deformed
nucleus. Let us assume that
interaction of a neutron with a nucleus is the sum
of the T-odd interaction, discussed above, and the strong one.
For evaluating the neutron-nucleus elastic scattering amplitude we'll
use the
Schroedinger equation (\ref {sr}).
The scattering amplitude at zero angle in the
third Born approximation is written as
\begin {eqnarray}
F (\vec k, \vec k) = -\frac {\varepsilon} {2\pi} \Biggl \{U
(\vec k, \vec k)~~~~~~~~~~~~~~~~~~~~~~~~~~~~~~
\nonumber
\\
+
2\varepsilon\int\frac {U (\vec k, \vec k ^\prime) U (\vec
k ^\prime, \vec k)}
{ k^2-k ^ {\prime 2} +i0} \frac {d^3\vec k ^\prime} {(2\pi) ^3}~~~~~
\nonumber
\\
+ (2\varepsilon) ^2\int\frac {U (\vec k, \vec k ^\prime) U
(\vec k ^\prime, \vec k ^ {\prime\prime}) U (\vec k ^
{\prime\prime}, \vec k)}
{ (k^2-k ^ {\prime 2} +i0) (k^2-k ^ {\prime\prime 2} +i0)}
\frac {d^3\vec k ^\prime} {(2\pi) ^3}
\frac {d^3\vec k^{\prime\prime}}{(2\pi)^3}
\nonumber
\\
+\cdots\Biggr\},
\label {born3}
\end {eqnarray}
where $U (\vec k ^\prime, \vec k) = \int
\mbox {e} ^ {-i\vec k ^\prime \vec r} V (\vec r) \mbox {e}
^ {i\vec k \vec r} d^3\vec r$ represents the Fourier
transform of the neutron-nucleus potential.
$U (\vec k ^\prime, \vec k)$ is the sum of
the strong interaction part (for simplicity we consider it is not 
to be
depending on the spin)
and  T-odd one:
\begin {eqnarray}
U (\vec k ^\prime, \vec k) =u_s (\vec k ^\prime-\vec
k) + \bigl (\boldsymbol \sigma\cdot (\vec k ^\prime + \vec k)
\times
( \vec k ^\prime
\nonumber
\\
-\vec k) \bigr) \bigl ((\vec k ^\prime + \vec k) \cdot
(\vec k ^\prime-\vec k) \bigr) u_T (\vec k
^\prime-\vec k).
\label {ufur}
\end {eqnarray}
For the  nucleus with the centre of symmetry $u_s (\vec k) =u_s
(-\vec k) $ and $u_T (\vec k) =u_T (-\vec k) $.
Substituting (\ref {ufur}) in (\ref {born3}) we find that the first and
second Born terms give zero contributions.
As a result, we have
\begin {eqnarray}
F_T (\vec k, \vec k) = -\frac {2\varepsilon^3} {\pi} \int \Biggl
\{
2\bigl (
\boldsymbol \sigma\times (\vec k ^\prime + \vec k) \cdot
(\vec k-\vec k^\prime ) \bigr)
\frac {1} {k^2-k ^ {\prime\prime 2}}
\nonumber
\\ \times
u_T (\vec k-\vec k ^\prime)
 u_s (\vec k ^\prime\vec -k ^ {\prime\prime}) u_s
(\vec k ^ {\prime\prime} -\vec k)
+
\bigl (\boldsymbol \sigma\times (\vec k ^\prime +\vec k ^
{\prime\prime})
\nonumber
\\
\cdot (\vec k ^\prime-\vec k ^ {\prime\prime})
\bigr)  \frac {((\vec k ^ {\prime\prime} + \vec k ^\prime)
(\vec k ^\prime-\vec k ^ {\prime\prime}))}
{ (k^2-k ^ {\prime 2}) (k^2-k ^ {\prime\prime 2})} u_s (\vec
k-\vec k ^\prime)
\nonumber
\\ \times
 u_T (\vec k ^\prime-\vec k ^
{\prime\prime}) u_s (\vec k ^ {\prime\prime} -\vec k)
\Biggr \}\frac {d^3\vec k ^\prime} {(2\pi) ^3}
\frac {d^3\vec k ^ {\prime\prime}} {(2\pi) ^3}. ~~~~~
\label{dev}
\end {eqnarray}
In the first item of (\ref{dev}) we change variables
$ \vec k ^\prime =\vec k +\vec q, ~ \vec k ^
{\prime\prime} = \vec k + \vec Q$, and in the second item
$ \vec k ^\prime = \vec k +\vec Q-\vec q, ~~
\vec k ^ {\prime\prime} = \vec k + \vec Q$ and come to
\begin {eqnarray}
F_T (\vec k, \vec k) =
-\frac {2\varepsilon^3} {\pi}
\int \Biggl \{-2 (\boldsymbol \sigma\times \vec k\cdot \vec q)
\biggl (
\frac {1} {k^2- (\vec k + \vec Q) ^2} \nonumber
\\
+ \frac {1} {k^2- (\vec k + \vec Q-\vec q) ^2} \biggr)
+2 (\boldsymbol \sigma\times \vec Q\cdot \vec q) \biggl (
\frac {1} {k^2- (\vec k +\vec Q) ^2} \nonumber
\\
-\frac {1} {k^2- (\vec k + \vec Q-\vec q) ^2} \biggr)
\Biggr \}
\nonumber
u_T (\vec q) u_s (\vec q-\vec Q) u_s (\vec Q)
\nonumber\\ \times
 \frac {d^3\vec q} {(2\pi) ^3}
\frac {d^3\vec Q} {(2\pi) ^3}
=
-\frac {8\varepsilon^3} {\pi} \int
( \boldsymbol \sigma\times \vec Q\cdot \vec q) \frac {1} {k^2-
(\vec k +\vec Q) ^2}
\nonumber \\ \times
u_T (\vec q) u_s (\vec
q-\vec Q)
u_s (\vec Q) \frac {d^3\vec q} {(2\pi) ^3}
\frac {d^3\vec Q} {(2\pi) ^3}.~~~~~
\label{ft1}
\end {eqnarray}
Deriving the last equality we have changed
the variables $ \vec Q =\vec Q ^\prime-\vec q ^\prime $,
$ \vec q =-\vec q ^\prime $ in terms containing the factor $ \frac {1} {k^2-
(\vec k +\vec Q-\vec q) ^2} $.
Using the formula
(\ref{telectr}) we express
$u_T (\vec k)$
through the Fourier transform of the charge distribution function
inside the nucleus $\rho (\vec k)$
(nucleus charge form factor):
\begin {equation}
u_T (\vec k) =\mu_T \frac {\pi Ze^2} {\varepsilon m^3}
 \frac{\rho(\vec k)} {k^2}.
\end {equation}
 In the rough approximation, being suitable
however for our purposes, the
strong interaction term can also be expressed through the Fourier transform
 of the nucleon distribution function in the nucleus
 and the nucleon-nucleon scattering amplitude
at zero
angle $f(0)$  :
\begin {equation}
u_s (\vec k) =- \frac {2\pi A f (0)} {\varepsilon} \rho (\vec k),
\label {siln}
\end {equation}
where
$A$ is the atomic number of the nucleus.
Thus, we assume,
that the charge distribution coincides with the matter density.
From (\ref{ft1})  we come to
\begin {eqnarray}
F_T(\vec k, \vec
k)=-\mu_T\frac{8Ze^2}{m^3}
(2\pi)^2A^2f^2(0)
\nonumber
\\
\times
\int
\frac {(\boldsymbol \sigma\times \vec Q\cdot \vec q)} {k^2-
(\vec k +\vec Q) ^2+i0} \frac {\rho (\vec q) \rho
(\vec q-\vec Q) \rho (\vec Q)} {q^2}
\nonumber
\\
\times
 \frac {d^3\boldsymbol
q} {(2\pi) ^3}
\frac {d^3\vec Q} {(2\pi) ^3}
\label {res1}
\end {eqnarray}
At high energies a simplification can be  achieved by using the
propagator in the eikonal approximation:
\begin {equation}
\frac {1} {k^2- (
\vec k +\vec Q) ^2+i0} \approx \frac {1} {-2 \vec  k\vec
Q+i0} = -P\frac {1} {2 \vec k\vec Q} -i\pi\delta (2\vec
k\vec Q).
\label {pp}
\end {equation}
The contribution of the first term of (\ref {pp}) vanishes as can be checked
by changing of  the variables
$ \vec Q =-\vec Q^\prime $,
$ \vec q =-\vec q ^\prime $ in the expression (\ref {res1}).
The deformed nucleus
form factor
can be taken in the form:
\begin {equation}
\rho (\vec q) = \mbox {e} ^ {-\beta q^2 +\beta ^\prime (\boldsymbol a \vec q)
^2}.
\label {furk}
\end {equation}
The unit vector $ \boldsymbol a $ is parallel to the axis of symmetry 
of the nucleus ($z$-axis)
and describes the orientation of the nucleus.
The expression (\ref {furk}) corresponds to the charge and matter 
distribution function:
\begin {eqnarray}
\wp (\vec r) = \frac {1} {(2\pi) ^3} \int \mbox {e} ^ {i\vec q\vec r} \rho (\vec q) d^3\vec q~~~~~~~~~~~~
 \nonumber
\\
 =\frac{1}{8\pi^{3/2}\beta\sqrt{\beta-\beta^\prime}}
\exp{\left(-\frac{r^2}{4\beta}-\frac{\beta^\prime z^2} {4\beta (\beta-\beta
^\prime)} \right)}.
\end {eqnarray}
\begin{figure*}
\vspace{-0 cm}
\hspace{2 cm}
\resizebox{0.7\textwidth}{!}{
  \includegraphics{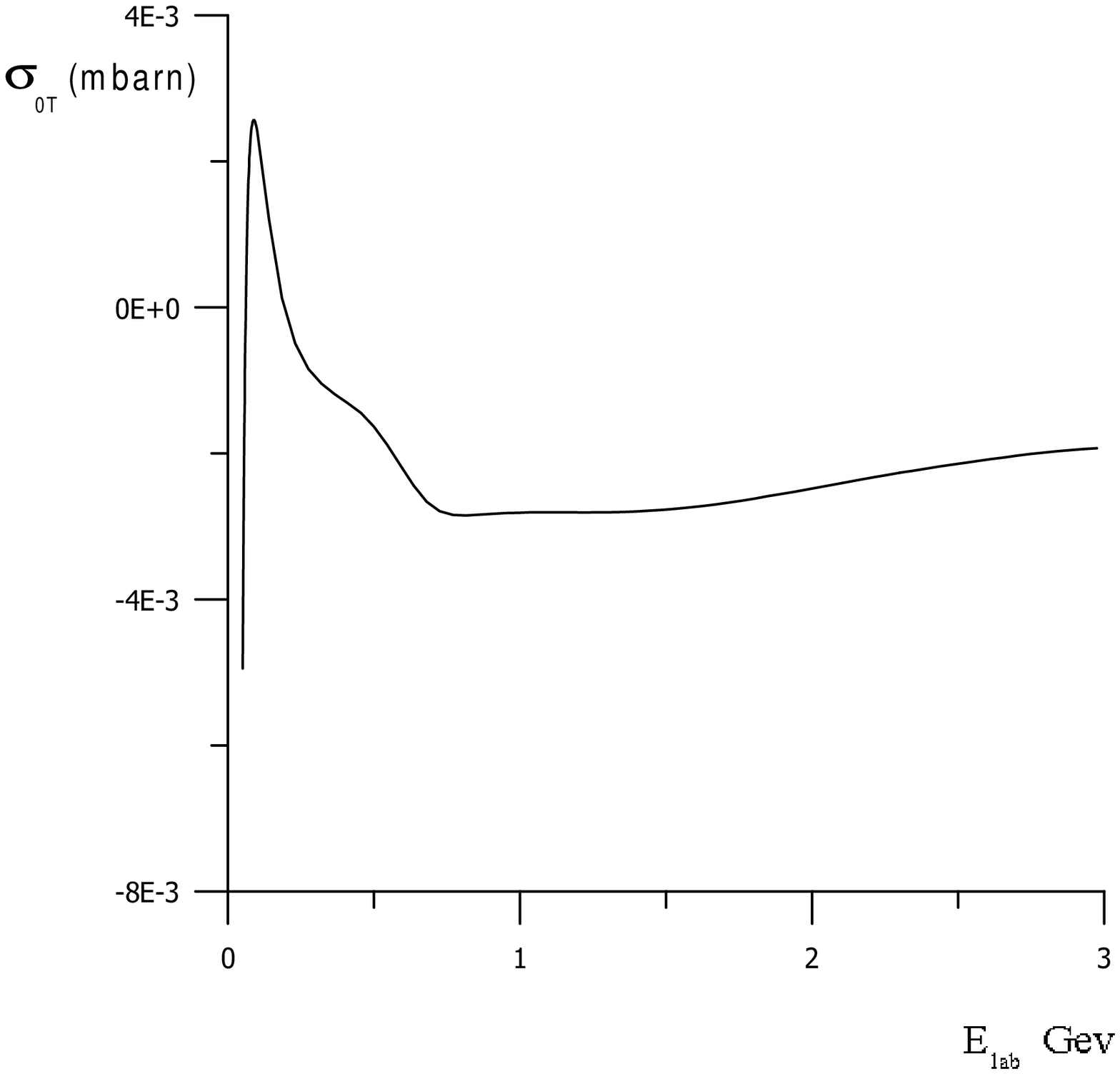}
}
\vspace*{-3. cm}
\caption{T-violating P-even cross section $\sigma_{\mbox{\tiny T}}
 =\sigma_{\mbox{\tiny 0T}} (\boldsymbol \sigma\cdot \vec S
\times {\vec k \over k}) (\vec S\cdot { \vec k \over k})$ for $^{165}\mbox{Ho}$ target at T-violating moment $\mu_T=1$.
}
\label{fig:2}       
\end{figure*}
$ \beta^\prime $ characterises the degree of 
deformation of the
nucleus
and is connected with the quadruple moment
\begin {equation}
{ \mathcal {Q}} =Z\int (3z^2-r^2) \wp (\vec r) d^3 \vec r =
-4 \, Z\beta ^\prime.
\end {equation}
For a small nucleus deformation  the nucleus root-mean-square radius is 
expressed
through $ \beta $:
\begin {equation}
R^2 =\int r^2\wp (r) d^3r\approx 6\beta.
\end {equation}
The calculation of the  integral
In the first order in
$ \beta ^\prime $  gives the following expression:
\begin {eqnarray}
\int \vec Q\times \vec q\frac {\delta (\vec k\vec
Q)} {q^2} \exp\bigl (-\beta q^2 +\beta ^\prime (\boldsymbol a\vec q)
^2-\beta (\vec q-\vec Q) ^2 \nonumber
\\
+
\beta ^\prime (\boldsymbol a\cdot (\vec q-\vec Q)) ^2 -\beta \vec Q^2
+\beta ^\prime (\boldsymbol a\vec Q) ^2\bigr) d^3\vec
Qd^3\vec q \nonumber
\\
\approx (\boldsymbol a\times \vec k) (\boldsymbol a\cdot \vec k)
\frac {\pi ^ {5/2}} {2k^3} \frac{\beta^\prime}{\beta^2}
\nonumber
\\ \times
\int_\beta^{\infty}
\frac{\beta-x}{(2x+\beta)^2(\beta+x)^{3/2}}dx.
\label{int1}
\end {eqnarray}
With the help of (\ref{int1}) we obtain the final formula:
\begin {eqnarray}
F_T (\vec k, \vec k) = i (\boldsymbol \sigma\cdot \boldsymbol a
\times \vec k) (\boldsymbol a\cdot \vec k)
\mu_T
 \nonumber
\\
\times
 \frac{Ze^2A^2f^2(0)}{8\sqrt{\pi}m^3k^3}\frac{\beta^\prime}{\beta^{7/2}} \int_1
^\infty \frac {(1-x) dx} {(2x+1) ^2 (1+x) ^ {3/2}}.
\end {eqnarray}
The unit vector $\boldsymbol a$ is expressed thorough the only
available
nucleus spin operator vector :
$\boldsymbol a={\vec S\over{\sqrt{S(S+1)}}}$.
Setting  T-violating moment
(expressed in $e/2m$ units)
$\mu_T=1$
we find that   T-odd cross
section  is about of $10^{-3}$ mbarn (Fig.\ref{fig:2})
for a $^{165}\mbox{Ho}$ ($S=7/2$) target.

It is possible to obtain the formula for the limiting case of
low energies, however, the  approximation (\ref {siln}) used for
the strong interaction becomes very rough.
At low energies the propagator can be approximated by:
\begin {equation}
\frac {1} {k^2- (
\vec k +\vec Q) ^2+i0} \approx-\frac {1} {Q^2} +
\frac {2\vec k\vec Q} {Q^4} -\frac {4 (\vec k\vec
Q) ^2} {Q^6}.
\end {equation}
The final formula at low energies looks like:
\begin {eqnarray}
F_T (\vec k, \vec k) = (\boldsymbol \sigma\cdot \boldsymbol a
\times \vec k) (\boldsymbol a\cdot \vec k)
 \mu_T
\nonumber
\\ \times
\frac{8Ze^2A^2f^2(0)}{15\pi m^3}
 \frac {\beta ^\prime}
{\beta^2}
\int_1 ^\infty \frac {(1-x) dx} {(x+1) ^3\sqrt {2x+1}}
\end {eqnarray}
 A magnitude of the amplitude is proportional to the
squared neutron wave number
$ k^2 $
at low energies,  whereas
at high energies it
decreases in inverse proportion
to
the wave number.
At $kR\sim 1$ both formulas give the same but overestimated order 
of the amplitude
magnitude. So, we restrict ourselves 
$\varepsilon-m=\mbox{E}_{\mbox{\tiny lab}}>50~~Mev$
(Fig.\ref{fig:2}),
where $kR>7$ and the
eikonal approximation should be valid.
\section{Estimates for T-odd magnetism for the other systems.}
It is of interest to study the consequences of T-violating P-conserving 
magnetism for
other systems.
\par
{\sl a) Electric dipole moment (EDM) of a neutron.}
The existing rigid experimental limit on
the  neutron EDM
($8\times10^{-26} e~ cm$)  allows one to obtain constraints on the T-odd P-even interactions.
Actually we have $P-odd~~T-odd=(P-even~~ T-odd)\times(P-odd ~~T-even)$. So,
P-conserving
beakdown of the time reversal symmetry
contributes to the neutron EDM
through  interference with P-odd weak
interaction.
\begin{figure}
\vspace{1.cm}
\hspace{0.5 cm}
\resizebox{0.4\textwidth}{!}{
  \includegraphics{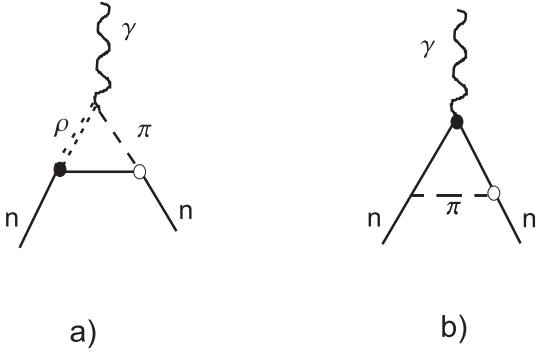}
}
\vspace{-0.cm}       
\caption{Diagrams contributing to the neutron EDM. Black circles denote
T-odd P-even vertexes, white circles denote P-odd T-even vertexes.
}
\label{dip}       
\end{figure}
The restriction $\bar g_\rho <10^{-3}$  on the relative
magnitude of the T-odd P-even nucleon-$\rho$-meson coupling
 has been obtained \cite{haxt} by calculating the Feynman graph in
Fig. \ref{dip}a.
The source of T-violation was 
the
$\rho$-meson-nucleon vertex and the source of
P-violation was the $\pi$-meson-neutron interaction.

We can consider the diagram in Fig.\ref{dip}b, corresponding  to the
T-odd magnetism contribution to the neutron EDM.
Both diagrams contain strong, electromagnetic and weak interaction vertices.
Hence they should give approximately the same restriction
on the relative strength of the T-violation, but, in the first case, 
T-violation occurs in strong
interaction, and in the second case it occurs in the electromagnetic one.
However,  due to the of-mass-shell character of the electromagnetic vertex
an additional suppression factor $\frac{(Pq)}{m^2}\sim \left( \frac{m_\pi}{m}\right)^2\approx 2\times 10^{-2}$
arises \cite{haff}. Thus the constraint on $\mu_T$  is expected to be $\mu_T\sim 10^{-1}$.
\par
{\sl b) positronium-like system decays.}
Let us now consider the positronium system.
The density of electrons (positrons) in positronium
state with total spin $J=1$ can be presented in the form \cite{cher}
\begin{eqnarray}
\rho (\vec r)=
A_0(r)+A_1\vec J\left( \boldsymbol \sigma _-+
\boldsymbol \sigma _+\right)
+A_2(\vec J\vec r)(\boldsymbol \sigma _-+
\boldsymbol \sigma _+)\cdot \vec r
\nonumber
\\
+\dots
+
T_0(\boldsymbol \sigma_- \times\boldsymbol \sigma_+\cdot\vec J)+
T_1((\boldsymbol \sigma _-\vec r)
(\boldsymbol \sigma _+\times \vec J\cdot \vec r)
\nonumber\\
-
 (\boldsymbol \sigma _+\vec r) (\boldsymbol
 \sigma _-\times \vec J\cdot \vec r))
+
T_2 ((\vec J\vec r) ((\boldsymbol
\sigma _- - \boldsymbol \sigma _+) \times \vec J
\cdot \vec r)
\nonumber\\+
((\boldsymbol
\sigma _- -\boldsymbol \sigma _+) \times \vec J
\cdot \vec r) (\vec J\vec r)),~~~~~~
\label{rho}
\end {eqnarray}
where $\vec r$ is the electron radius vector (the positron radius-vector 
is  $-\vec r$ ), $A_n, T_n$ are the
functions of $r$,
$\boldsymbol \sigma_-$, $\boldsymbol \sigma_+$ are the Pauli matrices of 
electron and positron, respectively.
The density is simultaneously the
spin density matrix of a positron and an
electron.
The positronium total spin operator  $\vec J$ is a parameter  describing the
positronium orientation. To find the density for the concrete positronium orientation
we must take the matrix element from (\ref{rho}) over a positronium
spin state i.e. replace $\vec J$ and $\vec J\otimes\vec J$ through the positronium
polarization $<\vec J>$ and quadropolarization $<\vec J\otimes\vec J>$ 
respectively.
The result of C-, P-, T- transformations on the density
can be described by the
operations
\begin{eqnarray}
C:\boldsymbol \sigma_+\rightarrow{\boldsymbol \sigma}_-,~~
\boldsymbol \sigma_-\rightarrow\boldsymbol \sigma_+,~~
 \vec r\rightarrow-\vec r~,
\nonumber
\\
T:\vec J\rightarrow-\vec J,~~ \boldsymbol\sigma_+\rightarrow-
{\boldsymbol \sigma}_+, ~~ \boldsymbol\sigma_-\rightarrow -\boldsymbol\sigma_-~,
\nonumber
\\
P: \vec r\rightarrow -\vec r.~~~~~~~~~~~~~~~~~~~~~~~~~~~~~~~~~~~~~
\end{eqnarray}
 We  can see that the terms proportional to $T_0, T_1, T_2$ are T-odd, C-odd, P-even terms.
But,
it is not possible to construct
T-odd P-even terms for the states with
$J=0$. From this fact we may conclude
that decays of positronium like systems with $1$ spin should be used 
to search indirect T- C-violation.

The terms  $T_0,\dots$ can originate from mixing of
the states with the same spatial parity but opposite charge parity due to
the P-even T-odd C-odd interaction
\begin{eqnarray}
{V}_T (\boldsymbol r) =- \mu_T\
\frac {3e^2} {2\varepsilon m^3}
  \bigl ((
\hat {\vec p} \vec r) \frac {1} {r^5} (
\vec r\cdot(\boldsymbol\sigma_- -\boldsymbol \sigma_+)\times \hat {\vec p})
\nonumber
\\
 + (
(\boldsymbol \sigma_--\boldsymbol \sigma_+)\times
\hat {\vec p} \cdot \vec r)
\frac {1} {r^5} (\vec r
\hat {\vec p}) \bigr) ~,
\end{eqnarray}
where $\mu_T$ is the T-violating electron moment
(the positron has the same)
and $m $ is the electron mass.
However for the orto-positronium state
$^3S_1~(J^{PC}=1^{- -})$ there is no the state with $J^{PC}=1^{-+}$ which 
can be mixed to it.
The charge parity of positronium is given by $C=(-1)^{l+s}$ and the spatial parity
is given by $P=(-1)^{l+1}$.
So T- C- violation can occurs only in the direct decay not under consideration here.
For the state $^1P_1~ (J^{PC}=1^{+-})$ there exists a state 
$^3P_1~(J^{PC}=1^{+ +})$ which can
mixed to it. The impurity can be estimated as
$\eta_T\sim \frac{V_T}{\Delta E}$,
where $V_T$ is the typical value of a T-odd interaction and $\Delta E $ 
is the splitting between
these levels. The
splitting $\Delta E$ can be produced by  tensor and spin-orbital interactions \cite{ber}
\begin{eqnarray}
V_s=\frac{3\alpha}{4m^2}\frac{1}{r^3}
\biggl(
\bigl(\vec r \times \vec p\cdot(\boldsymbol \sigma_-+\boldsymbol \sigma_+)\bigr)
\nonumber
\\
+\frac{(\boldsymbol \sigma_-\vec r)(\boldsymbol \sigma_+\vec r)}{r^2}-\frac{1}{3}(\boldsymbol \sigma_-\boldsymbol \sigma_+)
\biggr),
\end{eqnarray}
where $\alpha=e^2$ is the fine structure constant.
Typical values of an electron momentum and co-ordinate in positronium are
$p\sim m\alpha,~~r\sim \frac{1}{m\alpha}$ \cite{ber}.
Thus, we can estimate $V_T$ to be
\begin{equation}
V_T\sim \mu_T \frac{\alpha}{m^4}\frac{p^2}{r^3}\sim\mu_T m \alpha^6
\end{equation}
and
\begin{equation}
\Delta E\sim V_s\sim \frac{\alpha}{m^2}\frac{p}{r^2}\sim m \alpha ^4.
\end{equation}
As a result, for  the C- and T- odd impurity of the $^3P_1$ state 
to the $^1P_1$ state we have
\begin{equation}
\eta_T\sim\frac{V_T}{\Delta E}\sim\mu_T\alpha^2.
\end{equation}
For the branching ratio we get
\begin{equation}
\frac{\mathcal R (^1P_1\rightarrow  ^3S_1 +\gamma)}
{\mathcal R(^1P_1\rightarrow  ^3S_1+2\gamma)}
\sim\frac{\eta_T}{\alpha}\sim\mu_T \alpha.
\label{bran1}
\end{equation}
We take into account here that
the probability of
the decay into $^3S_1+2\gamma$ is reduced  by an additional
factor $\alpha$
compared to the decay to $^3S_1+\gamma$ \cite{ber}.
The measurement
of the branching ratio (\ref{bran1}) with the accuracy $10^{-2}$
gives
 constraint for electron's $\mu_T\sim 1$,
but it is far beyond the experimental possibilities of positronium
physics by now.

Let us consider a charmonium $c\bar c$ system, which is
similar to positronium. One gluon exchange
produces the Coulomb-like potential with a running constant
approximately equal to $\alpha_s=0.4$ \cite{led} 
(applicable also for the tensor interaction). Charmonium energy
levels can be described by this potential and a confinement
potential of oscillator type. The later is essential for  large
excitations and it
will not be  taken into account in our estimations.
Repeating our estimations for the present case we find:
\begin{eqnarray}
V_T\sim \mu_T \frac{\alpha}{m_c^4}\frac{p^2}{r^3}\sim\mu_T m_c \alpha\, \alpha_s^5,
\nonumber
\\
\Delta E\sim \frac{\alpha_s}{m_c^2}\frac{p}{r^2}\sim m_c \alpha_s ^4,
\nonumber
\\
\eta_T\sim\frac{V_T}{\Delta E}\sim \mu_T \alpha\,\alpha_s,
\end{eqnarray}
where $m_c$ is the mass of a charmed quark.
For the branching ratio of a $c\bar c$ system we find
\begin{equation}
\frac{\mathcal R (^1P_1\rightarrow J/\psi+\gamma)}
{\mathcal R(^1P_1\rightarrow J/\psi+2\gamma)}
\sim\frac{\eta_T}{\alpha}\sim\mu_T \alpha_s.
\end{equation}
The $^3P_1$ state of charmonium has experimentally been
identified
 and is called
$\chi_{c1}(1P)(3510)$ \cite{part2}. The $^1P_1$ state 
has not been clearly
identified by now.
A possible candidate would be $h_c(1P)(3526)$, but this needs 
confirmation \cite{part2}.

\begin{figure}
\vspace{0.cm}
\hspace{0.7 cm}
\resizebox{0.3\textwidth}{!}{
  \includegraphics{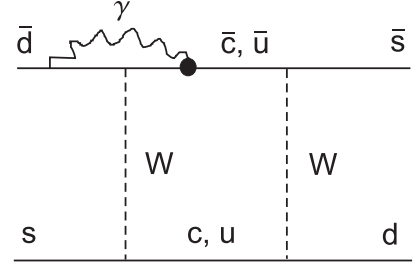}
}
\vspace{0.9cm}       
\caption{Diagram of T-odd radiation correction to the
$K^0-\bar K^0$ mixing. Black circle denotes T-odd P-even vertex.  }
\label{key}       
\end{figure}

{\sl c) Neutral kaon system.}
It is natural to assume that CP-violation is due to the Standard Model weak
interaction; however, another origin can not be excluded by now.
It is difficult to do some estimates
for T-odd magnetism for this case because of
competition of an
enhancement factor such as the small mass difference $m_{K_L}-m_{K_S}$ and
suppression factors such as the of-mass-shell character and 
spin-dependence of the T-odd P-even
vertex. One needs  to calculate radiation corrections
to the $K^0-\bar K^0$ mixing
with the T-odd
electromagnetic vertex (fig. \ref{key}).
A direct CP-violation can be estimated by evaluation of 
radiation
corrections similar to "penguin" \cite{led} diagrams.

\section{Conclusion}
Thus, we have
shown that besides P- T- odd
electric dipole moment the particle can have
T-violating P-conserving magnetic moment.
 We have considered the contribution of the T-odd magnetism to
the P-odd T-even neutron-nucleus forward elastic scattering amplitude.
We find, that the relative T-violation being of the order of unity 
corresponds to
the T-odd P-even cross section (Fig. \ref{fig:2}) being
about of $10^{-2}-10^{-3}~mbarn$ in the energy region of $50~MeV-3~GeV$.
The measurements  for 12 MeV neutrons
and $^{165}\mbox{Ho}$ target
give the constraint of
$10^{-2}~ mb$ on five fold correlation cross section \cite{huf0}.
If we relate this constraint to our
energy range we find that
$\mu_T\leq 1$.

It seems
neutron EDM
gives constraint $\mu_T\leq 0.1$.

Electrons and constituent quarks, in principle can
possess
T-violating P-conserving moments too. The way to search these may be the
observation of forbidden decay modes of positronium-like
systems from the $^3P_1$ and $^1P_1$ states.

\section{Acknowledgment}
The author is grateful to the Prof. V.G. Baryshevsky, Dr.  K. Batrakov 
and Dr. D.Matsykevich for discussions and remarks.

%
%


\begin{thebibliography}{}
\bibitem{for}
The most easier to deduce (\ref{sr}) from
Klein-Gordon equation:
$\biggl\{\left(\mbox{i}\frac{\partial}{\partial t}-V\right)^2+
\Delta -m^2\biggr\}\psi(\vec r,t)=0.$
Substituting $\psi(\vec r,t)=\mbox{e}^{-i\varepsilon t}\psi(\vec r)$ we find
$\bigl\{\left(\varepsilon^2-m^2\right)+\Delta\bigr\}=
\bigl\{2\varepsilon V-V^2\bigr\}\psi.$
So we can rely on the equation (\ref{sr}) takes into account
spin-less part of strong interaction correctly with accuracy $V/\varepsilon.$
\end{thebibliography}
\end{document}